# Indexing the Earth Mover's Distance Using Normal Distributions

Brian E. Ruttenberg
University of California, Santa Barbara
Santa Barbara, CA, USA
berutten@cs.ucsb.edu

Ambuj K. Singh
University of California, Santa Barbara
Santa Barbara, CA, USA
ambuj@cs.ucsb.edu

## ABSTRACT

Querying uncertain data sets (represented as probability distributions) presents many challenges due to the large amount of data involved and the difficulties comparing uncertainty between distributions. The Earth Mover's Distance (EMD) has increasingly been employed to compare uncertain data due to its ability to effectively capture the differences between two distributions. Computing the EMD entails finding a solution to the transportation problem, which is computationally intensive. In this paper, we propose a new lower bound to the EMD and an index structure to significantly improve the performance of EMD based K–nearest neighbor (K–NN) queries on uncertain databases.

We propose a new lower bound to the EMD that approximates the EMD on a projection vector. Each distribution is projected onto a vector and approximated by a normal distribution, as well as an accompanying error term. We then represent each normal as a point in a Hough transformed space. We then use the concept of stochastic dominance to implement an efficient index structure in the transformed space. We show that our method significantly decreases K–NN query time on uncertain databases. The index structure also scales well with database cardinality. It is well suited for heterogeneous data sets, helping to keep EMD based queries tractable as uncertain data sets become larger and more complex.

## 1. INTRODUCTION

Uncertain data is becoming more prevalent as the data generation capabilities of many scientific tools increases. The accuracy of many computational methods can be improved when the data uncertainty is retained, as little information is lost from the data acquisition phase to the analysis phase.

Uncertain data is frequently represented as probability distributions. Many traditional querying techniques suffer significant performance degradation when operating on distributions, due to the complexity of determining distance between uncertain objects. To address the issue of distance, the Earth Mover's Distance (EMD) has increasingly been utilized to query uncertain databases due to its ability to accurately retrieve similar distributions.

The EMD is a metric for computing the distance between two discrete probability distributions. The intuition behind the EMD is that it computes the minimum amount of "work" or "flow" required to transform one distribution into another. This property has made the EMD popular in recent years, finding use in image retrieval [15], cluster comparison [23] and shape matching [6].

The EMD is computationally expensive to derive, as the theoretical time complexity of the EMD is exponential in the number of distribution bins. Although empirically the complexity of the EMD is usually cubic in the number of bins [15], this high computation cost is still a significant bottleneck.

Nearest neighbor queries using the EMD require that the exact EMD between distributions must be computed, yet doing so on an entire database is prohibitive. Pruning objects based on EMD lower bounds has proven successful in reducing the time needed to answer nearest neighbor queries. Wichterich et al. [20] have shown that reducing the number of bins in the distributions can successfully prune candidate objects from the answer set. Recently, Xu et al. [21] have exploited the dual solution of the transportation problem to construct a B+ tree over a database of uncertain objects, which is then used to speed up query processing.

The above pruning methods suffer from some significant disadvantages. The reduction method proposed in [20] is implemented in the scan–and–refine architecture (SAR), and hence suffers from scalability shortcomings as an index structure is not implemented. The index proposed in [21] is constructed using pre–computed feasible solutions. The querying performance is contingent upon finding feasible solutions that are an accurate reflection of the underlying data set, and each feasible solution requires the use of two B+ trees. As the size of the database and the heterogeneity of the data increases, the number of B+ trees must also be increased in order to keep query times tractable. Hence these two methods suffer from scalability problems as the database cardinality and diversity is increased.

There is a clear need to develop a tight lower bound to the EMD that can be efficiently indexed and scaled to large datasets. Cohen et al. proposed a lower bound that projects distributions onto a vector, and computes the EMD in 1–dimension on the projection [4]. This bound is extremely tight and can be computed linearly in the number of bins. Figure 1 shows the average lower bound as a percentage of the actual EMD on a data set used in [20, 21]. The projection bound is highly accurate and a tighter bound than many other proposed lower bounds.

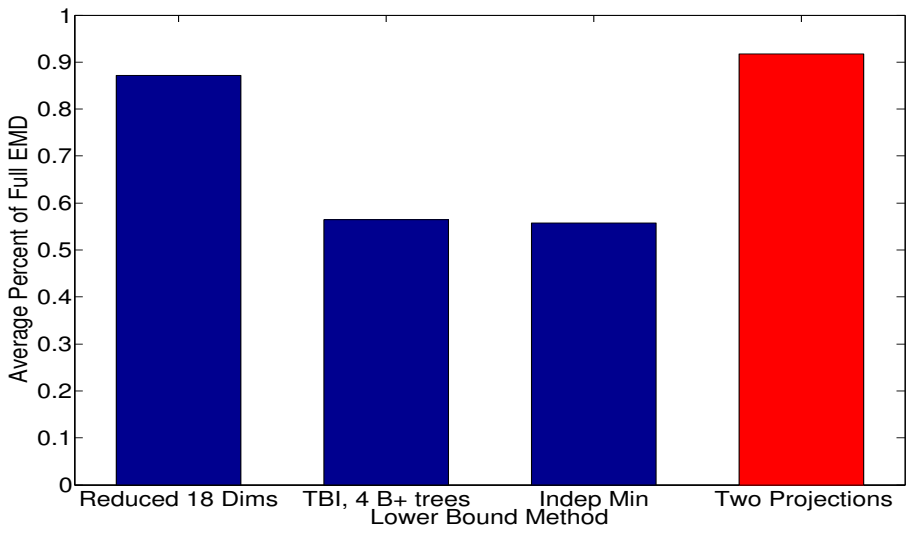

**Figure 1: Lower bounds on the RETINA data set. Each bar plots the average percent of the total EMD that the lower bound captures. The projection lower bound is a very tight lower bound.**

The projection bound [4] has not been utilized in K–NN queries due to the difficulty in indexing the bound. The projection bound is equivalent to the $L_1$ distance between the cumulative distribution functions (CDF) of the 1–dimensional discrete distributions. Indexing the projection bound is difficult due to the high dimensionality of the $L_1$ computation (the number of discrete distribution bins), coupled with the use of the $L_1$ distance that renders many dimensionality reduction methods ineffective. For example, the number of bins in the projection bound computation cannot be reduced using Fourier decomposition as there is no known equivalent of Parseval's theorem for the $L_1$ distance metric [13]. Hence, the development of a lower bound that is nearly as accurate as the projection bound but requires low $L_1$ dimensionality to compute would be extremely advantageous, as such a bound could be easily indexed using many common index structures.

In this paper, we propose a new lower bound to the EMD that approximates the projection lower bound. Our proposed bound combines an approximating normal distribution and an error term to compute a lower bound to the EMD in constant time. We show that this lower bound can be embedded into a novel low dimension index space and computed in $O(s)$ time, where $s$ is a small user defined parameter. We use the concept of stochastic dominance to successfully prune many potential nearest neighbor candidate objects. We test our method on several data sets, including a data set of over 600,000 distributions where our method is over twice as fast as the previous method. Our contributions can be summarized as follows:

- We develop a new lower bound to the EMD, called the *normal lower bound.*
- We show that the *normal lower bound* can be employed in a novel low dimension index that is constructed using the concept of stochastic dominance.
- We demonstrate that the *normal lower bound* significantly decreases K–NN query time, and scales well with large data sets.

## 2. RELATED WORK

The EMD is closely related to the family of mass transportation problems [14]. These problems are broadly concerned with the optimal movement of mass, flow or probability between two sets of data. The EMD was first introduced into the computer vision communities by Rubner et al. [15], though it has subsequently been shown to be an effective distance metric for many tasks, including comparing images [19] and shapes [18].

Numerous lower bounds to the EMD have been explored since it has been shown to be an effective metric for image retrieval and comparison [1, 2, 4, 11, 20]. Several papers have explored the use of the EMD in the scan–and–refine (SAR) querying architecture [1, 2, 20]. These methods focus on deriving accurate and efficient lower bounds to the EMD so some false candidates can be pruned away, though they suffer from scalability problems without the use of an index.

Xu et al. [21] has introduced a method that exploits the dual solution to the transportation problem to build a lower bound B+ tree index structure. This method, called TBI, creates several B+ trees based on a feasible solution to the EMD from a fixed data set. The B+ trees are then used at query time to eliminate candidates from a nearest neighbor query. The performance of the method is contingent upon finding feasible solutions that have high pruning power. Like the reduced EMD method, this may be unsatisfactory for a database that is changing frequently or composed of heterogeneous data. In addition, each feasible solution implemented requires two B+ trees, resulting in poor scalability as the diversity and size of the data set increases.

Indexing the normal lower bound is similar to the more general problem of indexing monotonic functions. There has been significant research on methods to index general functions, such as time series or probability distributions. For example, Keogh et. al. [9] and Yi and Faloutsos [22] propose approximating 1–dimensional time series data by adaptively dividing the time series into constant segments. They then build an index structure based on such constant segments. Ljosa and Singh [12] proposed a similar method to index expected distance functions or CDFs using non–constant line segments. These methods have serious shortcomings if applied to the EMD indexing problem. The methods proposed in [9] and [22] approximate arbitrary functions by a *constant* line segment, and hence require large amounts overhead to approximate a monotonic CDF. These indexing methods also focus on the $L_2$ distance. Many dimensionality reduction techniques that are successful for the $L_2$ distance do not provide any benefit for the $L_1$ distance [3].

The method in [12] is more suitable for approximating a CDF. However, this method was initially proposed to index the distance from a point to a CDF or expected distance function. Adapting the index to compute the $L_1$ distance between two CDFs would prove difficult, as each line segment in the approximation would have to be visited to compute the distance, due to the absolute value. Our method approximates CDFs by normal distribution CDFs, which are guaranteed to intersect at most once when the variances are unequal. In addition, there exists a closed form solution to the $L_1$ distance between two normal CDFs, enabling computation of the distance in constant time.

## 3. EARTH MOVER'S DISTANCE

The Earth Mover's Distance, as defined in [15], measures the minimum cost required to transform one histogram into another. While the formulation in [15] is generalized for histograms, we present a simple definition for distributions.

Given distribution weights $P = (p_1, \cdots p_n)$, distribution weights $Q = (q_1, \cdots q_n)$, and a set of distribution bin loca-

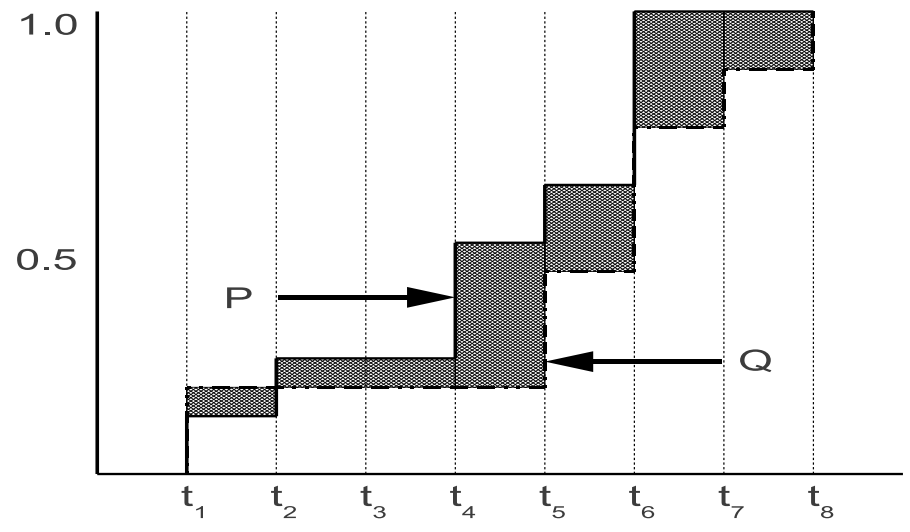

(a) 1–dimensional EMD

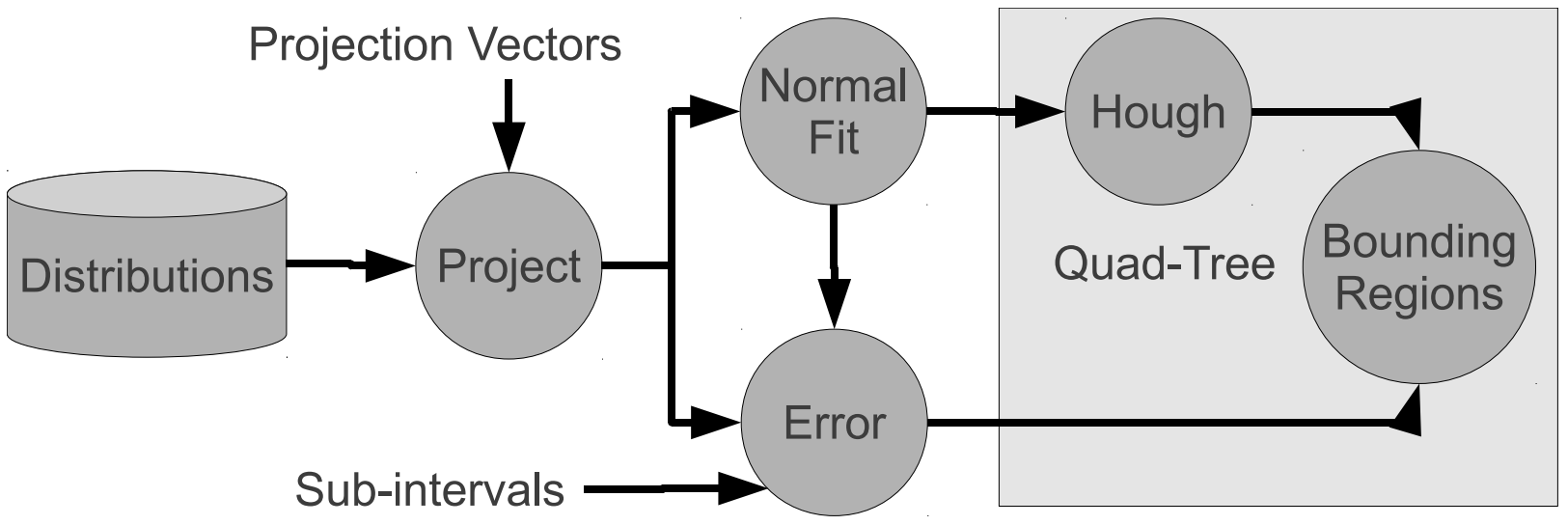

(b) Method Overview

**Figure 2: Figure 2(a): 1–dimensional EMD between projected distributions $P$ and $Q$. The EMD is the sum of the shaded areas of the figure, which is the absolute difference between the two distributions' CDFs. Figure 2(b): An overview of the normal lower bound index. Distributions are projected onto a vector. The distributions are fit to normal distributions and the errors are pre–computed. The normals are Hough transformed, and along with the errors, are used to create bounding regions in a quad–tree.**

tions $B = (b_1, \cdots b_n)$ that are common to both $P$ and $Q$, the total cost between $P$ and $Q$ is

$$F(P,Q) = \sum_{i=1}^{n}\sum_{j=1}^{n} f_{ij}c_{ij}$$

where $f_{ij}$ is the flow between bins $b_i$ and $b_j$, and $c_{ij}$ is the cost to move flow from $b_i$ to $b_j$. The choice of the cost is determined for each specific problem. For simplicity, we subsequently restrict our discussion to the $L_2$ norm, though the method can be applied to other cost structures. The EMD is then defined as

$$EMD(P,Q) = \min_F \ F(P,Q), \ \textit{subject to:}$$

$$f_{ij} \geq 0 \ \textit{and} \ \sum_{j=1}^{n} f_{ij} \leq p_i \ \textit{and} \ \sum_{i=1}^{n} f_{ij} \leq q_j$$

The EMD is the minimum cost needed to transform one distribution into another. The constraints ensure that the flow out of a bin is nonnegative and not more than the total weight in each bin.

The solution to the EMD attempts to minimize the distance that the weights $p_i$ must move to equal the weights $q_i$. When the distance between bins $b_i$ and $b_j$ is small, the solution will maximize the amount of flow between the bins; if large, it will minimize the amount of flow between the bins.

In general, solutions to the EMD use algorithms from linear programming, such as the transportation simplex [7]. The EMD has an empirically observed time complexity of approximately $O(n^3)$ [15], where $n$ is the number of bins in distribution. Hence, even small problem sizes can require significant time to compute.

## 3.1 EMD Projections

We briefly detail the EMD projection lower bound, as shown in [4]. If $S_j$ is a unit vector in $\mathbb{R}^d$, then

$$EMD(P,Q) \geq EMD(proj_{S_j}(P), proj_{S_j}(Q))$$

where $proj_{S_j}(P)$ is a projection of $P$ onto the vector $S_j$ (and similarly for $Q$). That is

$$proj_{S_j}(P) = (p_1 \ldots p_n, t_1, \ldots t_n), \ t_i = S_j^T \cdot b_i$$

where the weights are denoted as $p_i$, and the projected bins as $t_i$.

Furthermore, if $S = (S_1, \ldots S_{d'})$ is a set of orthogonal axes in $\mathbb{R}^d, d' \leq d$, then

$$EMD(P,Q) \geq \frac{1}{\sqrt{d'}} \sum_{j=1}^{d'} EMD(proj_{S_j}(P), proj_{S_j}(Q))$$

The EMD between $P$ and $Q$ is lower bounded by the sum of the EMDs on a set of orthogonal vectors, divided by the square root of the number of vectors. The EMD on a single projection can be computed in $O(n)$ time, a much more efficient computation than the full EMD. The following method of computing the EMD along a 1–dimensional projection is from [4].

THEOREM 1. *Let $C_{P,S_j}(t)$ denote the CDF of a discrete distribution $P$ projected onto $S_j$. Then given $C_{P,S_j}(t)$ and $C_{Q,S_j}(t)$, the EMD on the projection is*

$$EMD(proj_{S_j}(P), proj_{S_j}(Q)) = \int_{t_{min}}^{t_{max}} \left| C_{P,S_j}(t) - C_{Q,S_j}(t) \right| dt$$

*where $t_{min}$ and $t_{max}$ are the minimum and maximum projected bin values.*

PROOF. See [4], Theorem 4. □

Figure 2(a) shows the CDFs of two distributions on a projection. The EMD is the sum of the shaded areas in the figure, since that is exactly equivalent to the difference in the CDFs of the two distributions.

The computation of the EMD in this manner essentially amounts to the $L_1$ distance between two $n$–dimensional vectors. Note that this $L_1$ distance between the CDFs is not related to the cost to move flow between the original bins. Computing the projection EMD may be efficient compared to the cubic time requirement of the full EMD, but the $L_1$ distance between such large vectors is not easily indexed, due to the curse of dimensionality and the difficulty of dimensionality reduction for such a distance [3]. Thus, we introduce a new lower bound based on normal distributions that is easily indexed.

# 4. NORMAL LOWER BOUND

Figure 2(b) presents a simple overview of the normal lower bound and the subsequent indexing method. The bound utilizes the 1–dimensional projection EMD. There are two components to the normal lower bound; a distance between normal approximations to the projected distributions, and accompanying errors of the approximations. The normal approximations are subsequently transformed into points using Hough transformations, and combined with the error terms to create bounding regions in this transformed space. We detail construction of the index in Section 5. Throughout this section and next, we assume that each distribution is projected onto a single vector, hence we omit the projection subscript $S_j$. We will detail integration of multiple projection vectors in Section 5.3.

## 4.1 Normal Approximation

Given a 1–dimensional projected distribution $P$, we approximate $P$ by a normal distribution $\mathcal{N}(\mu_p, \sigma_p)$, where $\mu_p$ and $\sigma_p$ are the mean and variance of the projected distribution $P$.

*Definition 1.* **Normal CDF Integration** Let $N(\mu_p, \sigma_p)$ be a normal distribution, and let $\Phi_P$ be the CDF of the normal. Then the area under the CDF of the normal in the range of $[t_{min}, t_{max}]$ is defined as

$$\int_{t_{min}}^{t_{max}} \Phi_P(t)\,dt = t \cdot \Phi_P(t) + \phi_P(t) \Big|_{t_{min}}^{t_{max}}$$

where $\phi_P$ is the probability density function of a normal distribution.

Definition 1 is well known from the integration of the *error (erf) function*, which is equivalent to the CDF of the standard normal distribution. Despite the absolute value in Theorem 1, computing the EMD between 1–dimensional normal distributions does not require any numerical integration. As shown by Sinha and Zhou [17], two normal CDFs with unequal variances intersect at most once, and there is a closed form equation to determine the intersection point. Let $t_{is}$ be the intersection point between $\Phi_P$ and $\Phi_Q$. Then

$$t_{is} = \frac{\mu_p \sigma_q - \mu_q \sigma_p}{\sigma_q - \sigma_p} \qquad (1)$$

Given the intersection point and the range of integration $[t_{min}, t_{max}]$, we denote the normal EMD as $EMD_{\mathcal{N}}(\Phi_P, \Phi_Q)$, which is defined by rewriting Theorem 1 for normal distributions as

$$EMD_{\mathcal{N}}(\Phi_P, \Phi_Q) = \left| \int_{t_{min}}^{t_{is}} \Phi_P(t)\,dt - \int_{t_{min}}^{t_{is}} \Phi_Q(t)\,dt \right| + \left| \int_{t_{is}}^{t_{max}} \Phi_P(t)\,dt - \int_{t_{is}}^{t_{max}} \Phi_Q(t)\,dt \right|$$

If $t_{is}$ lies outside the range $[t_{min}, t_{max}]$, only one integration is performed over the entire range. Computing the normal EMD is a constant time operation due to the closed form of the normal CDF integration in Definition 1.

Next we briefly introduce the concept of stochastic dominance. This concept is integral to computing the lower bound, as well as for the subsequent indexing of the bound.

### 4.1.1 Stochastic Dominance

Stochastic dominance is the concept that in some fixed interval, the CDF of one distribution is always less than the CDF of another distribution [10]. Formally, we define stochastic dominance as

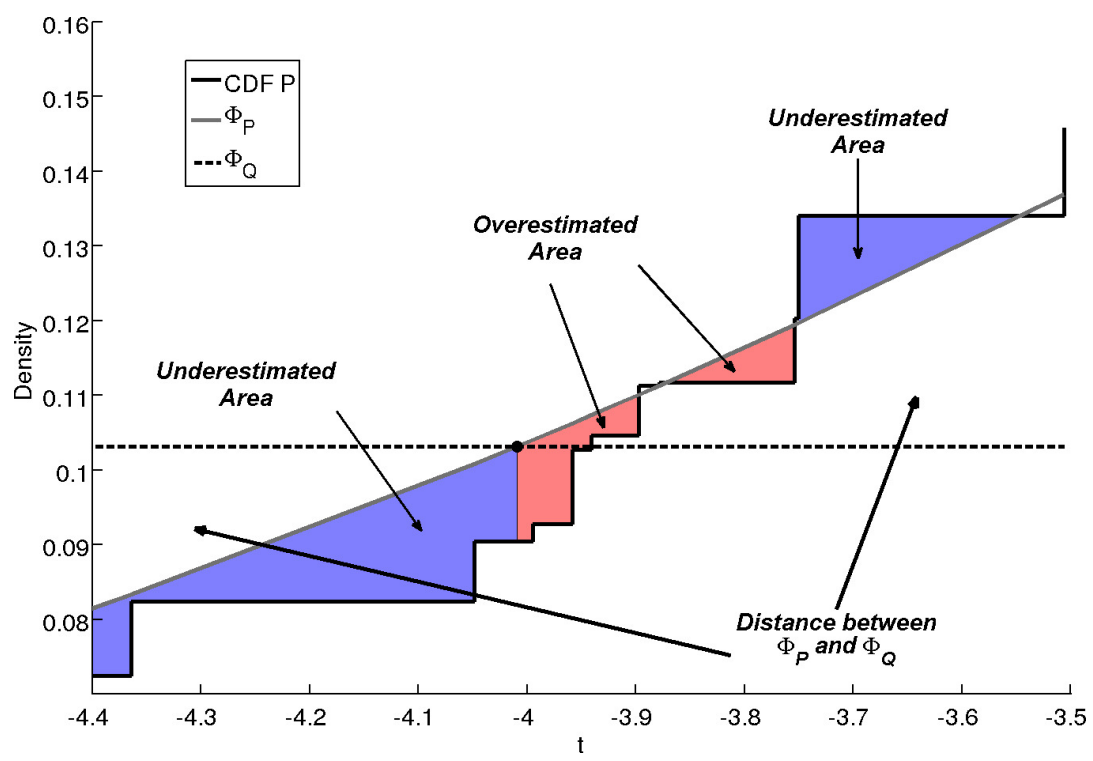

**Figure 3: The intersection of two normals and the CDF of one of the normals. The EMD between the normals has underestimated the actual projected EMD by the area in blue, and overestimated the EMD by the amount in red. This scenario assumes that $Err_Q$ is zero.**

*Definition 2.* **Stochastic Dominance.** A CDF $C_p$ stochastically dominates another CDF $C_q$ if

$$C_p(t) < C_q(t) \; \forall\, t \in [t_{min}, t_{max}]$$

If $C_p$ dominates $C_q$, we write $C_p \prec C_q$.

Note that varying definitions of stochastic dominance exist, but we use this definition in this work. Observe that two distributions may not dominate each other. In such a scenario, the two CDFs must intersect at least once, and in the case of normal CDFs, exactly once. We define stochastic dominance only in a fixed range $[t_{min}, t_{max}]$, as outside this interval the dominance property is not guaranteed.

Stochastic dominance provides several properties that assist in computation of the normal lower bound. First, if $\Phi_P \prec \Phi_Q$ in $[t_{min}, t_{max}]$, then we know that the integration of $\Phi_P$ is less than $\Phi_Q$ in the same range. In addition, if $\Phi_P$ intersects $\Phi_Q$ in the range $[t_{min}, t_{max}]$, then we know that

$$\Phi_P \prec \Phi_Q \text{ over } [t_{min}, t_{is}]$$
$$\Phi_Q \prec \Phi_P \text{ over } [t_{is}, t_{max}]$$

is true, or the converse is true, where $t_{is}$ is the intersection point between $\Phi_P$ and $\Phi_Q$.

## 4.2 Error Compensation

In order to take advantage of the constant time distance computation between normals, we must also compensate for the error incurred when each CDF is fit to a normal.

*Definition 3.* **Normal Approximation Error.** For a normal CDF $\Phi_P$, we define the normal approximation error $Err_P(t)$ incurred at point $t$ as

$$Err_P(t) = C_P(t) - \Phi_P(t) \qquad (2)$$

If at point $t$ the CDF of $P$ is greater than the normal approximation of $P$, we have a positive error, and a negative error in the reverse scenario.

Unfortunately, the errors cannot simply be accumulated over the entire distribution range and combined with the normal EMD to produce the lower bound. This is a result of the intersection point between two distributions impacting how the errors compensate for the normal approximation.

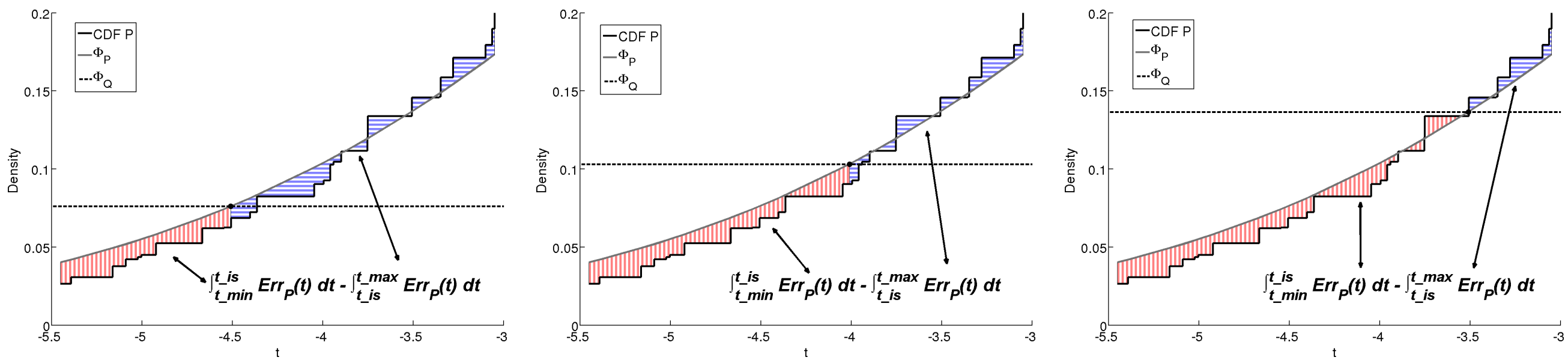

(a) (b) (c)

**Figure 4: Three different intersection points in a sub–interval. The difference between the errors before and after an intersection point is needed to compute the normal lower bound. Hence these values are pre–computed for a sub–interval, and the minimum and maximum differences are recorded.**

Consider the example in Figure 3, which shows two normal approximations $\Phi_P$ and $\Phi_Q$, and the CDF $C_p$. Let us assume that $C_q$ is exactly normal, meaning $Err_Q$ is zero everywhere. Before the intersection point of the normal CDFs, $\Phi_P$ dominates $\Phi_Q$. Therefore, the normal EMD between $\Phi_P$ and $\Phi_Q$ is an *underestimate* of the projected EMD by the amount of negative error in $Err_P$ before the intersection, shown in blue. Conversely, after the intersection point, the dominating relationship has changed. Therefore, the negative error is now the amount that the normal EMD has *overestimated* the projected EMD, shown in red.

Observe in the figure that after the intersection, the positive and negative errors will cancel each other out, as the positive error is an underestimate while the negative error is an overestimate. The entire error after the intersection point can be summed together and will be close to zero, since the positive and negative areas after the intersection are roughly equal. This means that for the example in the figure, the projected EMD is at least

$$EMD_{\mathcal{N}}(\Phi_P, \Phi_Q) - \int_{t_{min}}^{t_{is}} Err_P(t)\,dt + \int_{t_{is}}^{t_{max}} Err_P(t)\,dt \quad (3)$$

Note that if $\Phi_Q$ dominated $\Phi_P$ before the intersection, we would reverse the signs on the error terms.

We cannot pre–compute the error integrals in Eqn. (3) to compute a lower bound to the projected EMD at query time. The integrals in the equation depend upon the intersection, and obviously we do not know in advance the intersection point between a query and database object. However, in lieu of knowing the actual intersection point from a query, we break the interval $[t_{min}, t_{max}]$ into $s$ sub–intervals, and compute a pessimistic estimate of the error summations in each sub–interval.

We visit each potential intersection point in a sub–interval and pre–compute the worst case error terms. For example, Figure 4 shows one sub–interval with three different potential intersection points. At each intersection point the difference between the total error before the intersection and after the intersection is computed, as in Eqn. (3). In each figure, the error in the blue region is accumulated and subtracted from the accumulation of the error in the red region. We then store the minimum and maximum of these values, as we do not yet know the sign on the error terms until query time (if the error is subtracted we will need the maximum, and the minimum if the error is eventually added). The only intersection points we need to consider are where the discrete CDF changes or intersects its normal approximation, so there are a finite amount of intersection points that need to be checked.

The minimum and maximum error differences are pre–computed for all $s$ sub–intervals, which we denote as $Err_{min,P}$ and $Err_{max,P}$. These error differences can then be easily retrieved via a constant lookup at query time and either added or subtracted to the normal EMD, depending on how the two normals intersect. We perform the same calculation of the minimum and maximum error for each query as well (at query time, of course).

Formally, we define $Err_{min,P}$ as follows

$$Err_{min,P}(t_j) = \begin{cases} \min\limits_{t_{is}\in[s_i,s_{i+1}]} \{\int_{t_{min}}^{t_{is}} Err_P(t)\,dt - \int_{t_{is}}^{t_{max}} Err_P(t)\,dt\} \\ \qquad \text{if } s_i < t_j \leq s_{i+1} \\ \int_{t_{min}}^{t_{max}} Err_P(t)\,dt \quad \text{if } t_j \leq t_{min} \bigvee t_j > t_{max} \end{cases} \quad (4)$$

with $Err_{max,P}(t_j)$ defined in the same manner but with a maximization instead of a minimization.

## 4.3 Computing the Normal Lower Bound

Using stochastic dominance, we can now compute the normal lower bound between two distributions on a projection. We denote the normal lower bound $EMD_{LB}(\Phi_P, \Phi_Q)$ in the range $[t_{min}, t_{max}]$ as

$$EMD_{LB}(\Phi_P, \Phi_Q) = \begin{cases} EMD_{\mathcal{N}}(\Phi_P, \Phi_Q) + Err_{min,P}(t_{is}) - Err_{max,Q}(t_{is}) \\ \quad \text{if } \Phi_Q \prec \Phi_P \bigvee \sigma_p > \sigma_q \\ EMD_{\mathcal{N}}(\Phi_P, \Phi_Q) - Err_{max,P}(t_{is}) + Err_{min,Q}(t_{is}) \\ \quad \text{if } \Phi_P \prec \Phi_Q \bigvee \sigma_p < \sigma_q \end{cases}$$

where $t_{is}$ is the intersection point between $\Phi_P$ and $\Phi_Q$.

THEOREM 2.

$$EMD_{LB}(\Phi_P, \Phi_Q) \leq EMD(proj_{S_j}(P), proj_{S_j}(Q))$$

PROOF. See Appendix A. □

Computing the normal lower bound is a constant time operation. With a given query, the EMD between the normals is computed using the closed form definite integral. The intersection point is then computed using Eqn. (1), and the pre–computed error terms are retrieved and the bound is then computed.

# 5. INDEXING THE LOWER BOUND

The normal lower bound is a tight lower bound to the EMD. However, despite the $O(1)$ complexity of the bound, computing the lower bound on a large database of objects can still be very time consuming. An index structure based on the normal lower bound would be desirable to handle scalability of the data.

There are several challenges with the bound that must be resolved in order to build an effective index on this distance metric. First, we must define the space and structure of the index. Meaning, we must determine how each projected distribution will be represented in the index space. Second, we must ensure that computing the normal lower bound in the index space is still efficient. We now detail how we resolve both of these challenges.

## 5.1 Dominance Space

Normal distributions have the desirable property that stochastic dominance is preserved in lines that are computed using the mean and variance of the distributions.

THEOREM 3. *For* $t \in [t_{min}, t_{max}]$,

$$\Phi_P(t) \prec \Phi_Q(t) \;\Leftrightarrow\; \frac{t-\mu_p}{\sigma_p} \prec \frac{t-\mu_q}{\sigma_q}$$

PROOF. As shown in [17], $\Phi(\frac{t-\mu_p}{\sigma_p})$ is less than $\Phi(\frac{t-\mu_q}{\sigma_q})$ (where $\Phi$ is the standard normal CDF) if and only if $\frac{t-\mu_p}{\sigma_p} < \frac{t-\mu_q}{\sigma_q}$. The definition of dominance states that function $A$ must be strictly less than function $B$ in the given interval. Thus $\frac{t-\mu_p}{\sigma_p} < \frac{t-\mu_q}{\sigma_q}$ satisfies the definition of dominance and the theorem is true. □

In order to take advantage of this property we perform a Hough transformation [8] on the normal CDFs. Hough transformations convert line segments into points in the parameter space of the line segments. Using the standard y–intercept form of a line, $y = m \cdot t + b$, each normal is represented by a slope ($m$) and a y–intercept ($b$). We rewrite $\frac{t-\mu_p}{\sigma_p}$ into the standard y–intercept equation for line. We set $m = \frac{1}{\sigma}$ and $b = \frac{-\mu}{\sigma}$. Each normal CDF $\Phi_P$ is represented as a tuple $(m_p, b_p)$ in this transformed space. We term this transformed space *dominance space*, because it is easy to define dominance relationships in geometric terms.

Consider Figure 5(a), where we plot $y = m{\cdot}t{+}b$ for several normals. We denote the blue line as the line for $\Phi_P$, which dominates all the black lines. Additionally, the red dotted line has no dominating relationship with any of the other normals. We observe that at $t_{min}$ (the smallest $t$ value) the dominance relationship is preserved. That is,

$$\frac{t-\mu_p}{\sigma_p} \prec \frac{t-\mu_q}{\sigma_q} \rightarrow m_p \cdot t_{min} + b_p < m_q \cdot t_{min} + b_q \quad (5)$$

for some $\Phi_Q$ that is dominated by $\Phi_P$. There exists a region in dominance space such that at $t_{min}$, the normals defined by points in the region are always greater than $\Phi_P$ at $t_{min}$. This region can easily be derived from Eqn. (5) as

$$\begin{aligned} m_p{\cdot}t_{min} + b_p &< m_q{\cdot}t_{min} + b_q,\ \forall\, m_q, b_q \\ (m_p{\cdot}t_{min} + b_p) - m_q{\cdot}t_{min} &< b_q,\ \forall\, m_q \\ -t_{min}{\cdot}m_q + (m_p{\cdot}t_{min} + b_p) &< b_q,\ \forall\, m_q \end{aligned} \quad (6)$$

That is, we have defined a line in *dominance space*, with a slope of $-t_{min}$ and a y–intercept of $(m_p \cdot t_{min} + b_p)$, such that all normals with a $b_q$ value above the line are greater than $\Phi_P$ at $t_{min}$. This can be seen in Figure 5(b) as the dotted line. The points in dominance space for the black lines in Figure 5(a) are all above the dotted line in Figure 5(b) because at $t_{min}$, they are all greater than $P$.

Similarly, we can find another region using $t_{max}$, as shown in Figure 5(c). Note how the red line from Figure 5(a) that intersected all the other lines is *not* above this new dotted line, as at $t_{max}$, it is less than $\Phi_P$. We denote these lines in dominance space as dominance lines.

From Theorem 3 and Eqn. (6) we get a corollary regarding the dominance lines.

COROLLARY 1. *For* $t \in [t_{min}, t_{max}]$,

$$\frac{t-\mu_p}{\sigma_p} \prec \frac{t-\mu_q}{\sigma_q} \Leftrightarrow \begin{cases} b_q > -t_{min} \cdot m_q + (m_p \cdot t_{min} + b_p) \\ \textit{and} \\ b_q > -t_{max} \cdot m_q + (m_p \cdot t_{max} + b_p) \end{cases}$$

Corr. 1 implies that given a point $(m_p, b_p)$ in dominance space, there exists a region where $\Phi_P$ dominates all other normals. Conversely, if we switch the inequalities in Corr. 1, we can determine the region where $\Phi_P$ is dominated by all normals, which is the area below the intersection of the two lines in Figure 5(c). Any points in dominance space not contained in these regions must intersect $\Phi_P$ somewhere between $t_{min}$ and $t_{max}$.

Observe that the slope of the dominance lines are $-t_{min}$ and $-t_{max}$. As $t_{min}$ and $t_{max}$ are fixed for the database (since they are based on the projection), *all* dominance lines have the same slope, and thus are parallel. The shape of this region is the same no matter where the point $(m_p, b_p)$ of interest is placed.

Finally, we state the following theorem about dominance space.

THEOREM 4. *Denote the region that* $(m_p, b_p)$ *dominates as R, so that* $\Phi_P \prec \Phi_r\, \forall r \in R$. *Then for all points* $(m_q, b_q)$ *where* $\Phi_Q \prec \Phi_P$,

$$EMD_{\mathcal{N}}(\Phi_P, \Phi_Q) \le EMD_{\mathcal{N}}(\Phi_r, \Phi_Q),\ \forall r \in R \quad (7)$$

PROOF. This follows from the definition of dominance. If $\Phi_P \prec \Phi_r$, then $\Phi_P$ must be less than $\Phi_r$ at every point in the range, and hence the area under the CDF for $\Phi_P$ must be less than $\Phi_r$. Since $\Phi_Q \prec \Phi_P$, the area under $\Phi_Q$ must also be less than $\Phi_P$, and hence $EMD_{\mathcal{N}}(\Phi_P, \Phi_Q)$ must be less than any $EMD_{\mathcal{N}}(\Phi_r, \Phi_Q)$. □

The index is implemented in dominance space, and each projected distribution is represented as a point in this space. Next we detail how dominance lines and Theorem 4 are utilized to lower bound a set of points in dominance space.

## 5.2 Bounding Regions

An efficient index structure needs to have the capability of pruning large amounts of potential candidates from a K–NN query without resorting to computation of the lower bound between a query and each database object. This can be accomplished by creating bounding regions around points in dominance space, and computing the lower bound between a query and a bounding region. If the lower bound between the query and the bounding region is too large to be considered as a nearest neighbor, then all the points within the bounding region can be pruned with a single computation. The bounding regions that are utilized in the index are based on dominance lines.

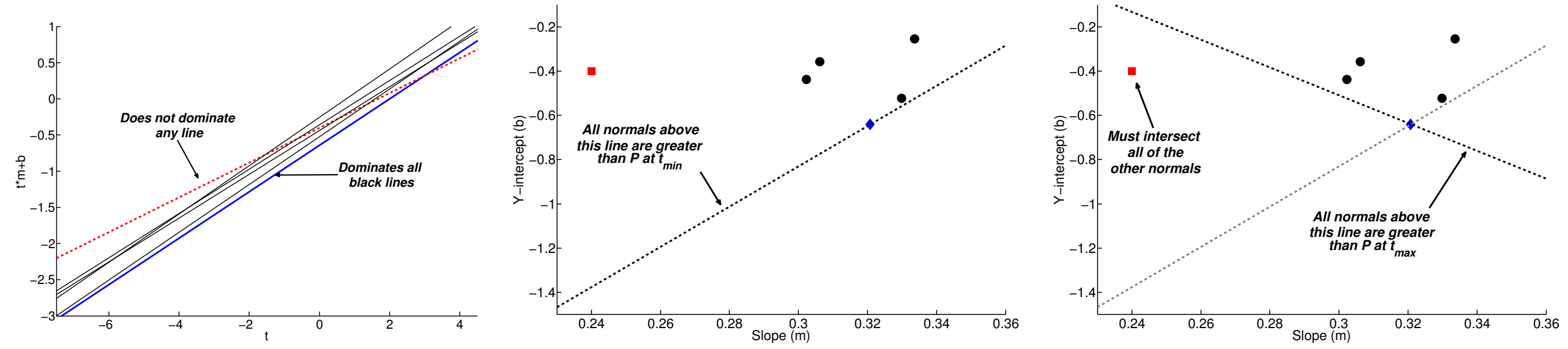

(a) (b) (c)

**Figure 5: Finding dominance lines in dominance space. At the endpoints of the lines in Figure 5(a), the dominance property must hold. All points in dominance space that have their endpoints dominated by the blue line must be above the two dominating lines in dominance space. Note that the red point has only one endpoint dominated by the blue line, hence it is only above one of the dominating lines in Figure 5(c).**

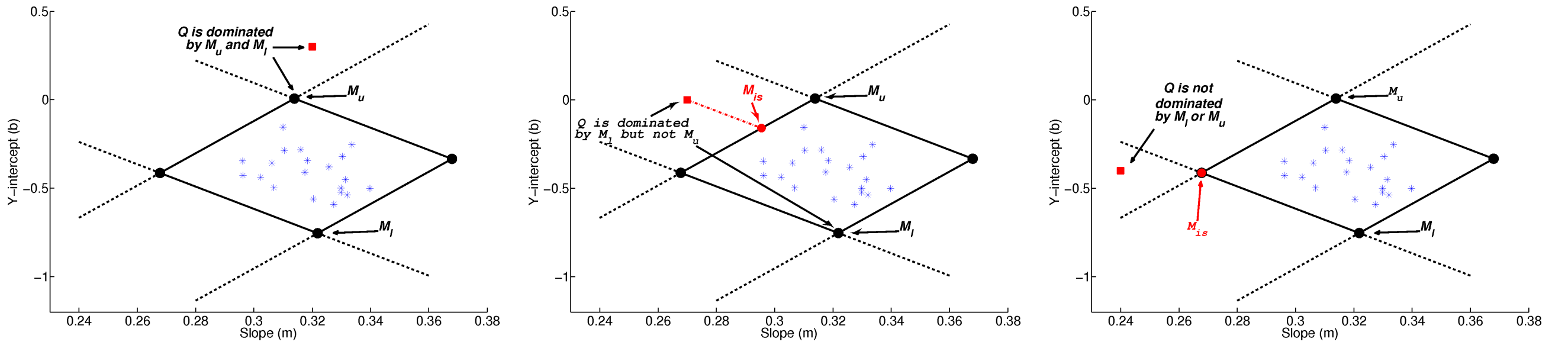

(a) Complete Dominance (b) Partial Dominance (c) No Dominance

**Figure 6: A bounding region in dominance space, and examples of the three scenarios where the distance from a point $Q$ to a bounding region must be considered. In Figure 6(a), $Q$ is completely dominated by every point in $M$. In Figure 6(b), $Q$ is partially dominated by $M$, and in Figure 6(c), $Q$ has no dominating relationship with $M$ (it must intersect every point in $M$). Note that every point $Q$ must fall into one of these scenarios, though the positions in dominance space may be different than depicted in the figures.**

*Definition 4.* **Bounding Region.** Given a set of points in dominance space, we define the bounding region (BR) $M$ of the points by two points $M_l$ and $M_u$ in dominance space, where

$$M_l = (m_l, b_l) : \Phi_{M_l} \prec \Phi_{m_i} \; \forall \, m_i \in M$$
$$M_u = (m_u, b_u) : \Phi_{m_i} \prec \Phi_{M_u} \; \forall \, m_i \in M$$

Recall from the previous section that the slope of the dominating lines are solely defined by $[t_{min}, t_{max}]$. As a result, only two points are needed to define a bounding region, since the other two points can be derived from the intersection of the dominating lines emanating from $M_l$ and $M_u$. This is similar to Euclidean space where only two points are needed to define a rectangle.

For the moment, we assume that $t_{min} < 0$ and $t_{max} > 0$. In such a scenario, $M$ forms a diamond shaped bounding region in dominance space. Figure 6(a) shows the dominating region for a series of points shown in blue. The bounding region is the area enclosed by the solid black lines, with $M_u$ being top of the diamond region, and $M_l$ the bottom.

We can easily compute the normal lower bound between a query $Q$ and all $m_i \in M$. However, in order to prune all the points in $M$, we need to compute the normal lower bound between $Q$ and the BR $M$. Unfortunately, computing a lower bound to a bounding region is not as simple to compute as the normal lower bound.

First, we define the minimum and maximum error differences for a BR similar to a point as

$$Err_{min,M}(t_{is}) = \min_{m_i \in M} Err_{min,m_i}(t_{is})$$
$$Err_{max,M}(t_{is}) = \max_{m_i \in M} Err_{max,m_i}(t_{is})$$

Meaning, the errors of some sub–interval $s_i$ for a bounding region are just the minimum and maximum of the sub–interval $s_i$ for any point within $M$.

We denote the normal lower bound specifically for bounding regions as $EMD_{BR}(M, \Phi_Q)$. We must show that

$$EMD_{BR}(M, \Phi_Q) \leq EMD_{LB}(\Phi_{m_i}, \Phi_Q) \; \forall \, m_i \in M$$

in order to prune a bounding region from a candidate set.

There are three different ways that $EMD_{BR}(M, \Phi_Q)$ is computed, depending on where the query point is relative to $M$. An example of each scenario is shown in Figures 6(a), 6(b) and 6(c). Similar to the normal lower bound between points, the lower bound for bounding regions is composed of a normal EMD and error terms. The normal EMD term for $EMD_{BR}(M, \Phi_Q)$ is the minimum normal EMD between $Q$ and any point in $M$. Determining this minimum distance point greatly depends on where $Q$ is relative to $M$, hence the three different ways that $EMD_{BR}(M, \Phi_Q)$ is computed.

### 5.2.1 Complete Dominance

The complete dominance scenario is shown in Figure 6(a). In this scenario, $Q$ is completely dominated by the entire BR.

From Theorem 4, we know

$$\Phi_{M_u} \prec \Phi_Q \Rightarrow$$
$$EMD_{\mathcal{N}}(\Phi_{M_u}, \Phi_Q) \leq EMD_{\mathcal{N}}(\Phi_{m_i}, \Phi_Q) \; \forall \; m_i \in M$$

$M_u$ is the point in $M$ that has the minimum normal distance to $Q$. $EMD_{BR}(M, \Phi_Q)$ is then computed as

$$EMD_{BR}(M, \Phi_Q) =$$
$$EMD_{\mathcal{N}}(\Phi_{M_u}, \Phi_Q) - Err_{max,M}(t_{is}) + Err_{min,Q}(t_{is})$$

Note that if $Q \prec M_l$, we use $M_l$ instead of $M_u$ and the appropriate minimum or maximum errors.

### 5.2.2 Partial Dominance

The partial dominance scenario is depicted in Figure 6(b). In this case, the query point $Q$ is dominated by $M_l$ but not $M_u$, and as a result, we cannot use Theorem 4 to find the minimum distance normal.

Fortunately, since the EMD is a metric, we can use the triangle inequality to lower bound the minimum normal EMD from $Q$ to $M$. We denote the intersection point of the dominating line from $Q$ with $M$ as $M_{is}$, shown as the red point in Figure 6(b). We know that the minimum distance point must lie above the intersection of the dominance line from $Q$ with $M$ (red in the figure), as follows from the complete dominance scenario in the previous subsection.

We define $EMD_{BR}(M, \Phi_Q)$ in the partial dominance scenario as

$$\begin{aligned} EMD_{BR}(M, \Phi_Q) = \\ \frac{1}{2}\,[EMD_{\mathcal{N}}(\Phi_{M_u}, \Phi_Q) + EMD_{\mathcal{N}}(\Phi_{M_{is}}, \Phi_Q) \\ -EMD_{\mathcal{N}}(\Phi_{M_{is}}, \Phi_{M_u})] - \max_{s_i \in s}\{Err_{max,M}(s_i)\} \\ + \min_{s_i \in s}\{Err_{min,Q}(s_i)\} \end{aligned} \tag{8}$$

The triangle inequality is used to bound the minimum normal distance, and the maximum and minimum errors over all sub–intervals from $M$ and $Q$ are used to ensure that the error terms are always less than any errors from a point inside $M$. If $Q \prec M_u$ but not $M_l$, then we use $M_l$ instead of $M_u$ and the appropriate minimum or maximum errors.

THEOREM 5. *If a query point $Q$ is dominated by $M_l$ but not $M_u$, then*

$$EMD_{BR}(M, \Phi_Q) \leq EMD_{LB}(\Phi_{m_i}, \Phi_Q) \, \forall \, m_i \in M$$

*where $EMD_{BR}(M, \Phi_Q)$ is computed as defined in Eqn. (8).*

PROOF. See Appendix (B) □

### 5.2.3 No Dominance

The last scenario is depicted in Figure 6(c). In this case, $Q$ has no dominance relationship with any points in $M$. This case is just an extension of partial dominance. We denote $M_{is}$ as the intersection point between the dominance lines from $M_l$ and $M_u$, as depicted in red on the figure. We use the triangle inequality to bound the minimum distance normal using $M_l$ and $M_{is}$, then repeat using $M_u$ and $M_{is}$, and take the minimum value. The rest of the terms in $EMD_{BR}(M, \Phi_Q)$ for partial dominance are the same.

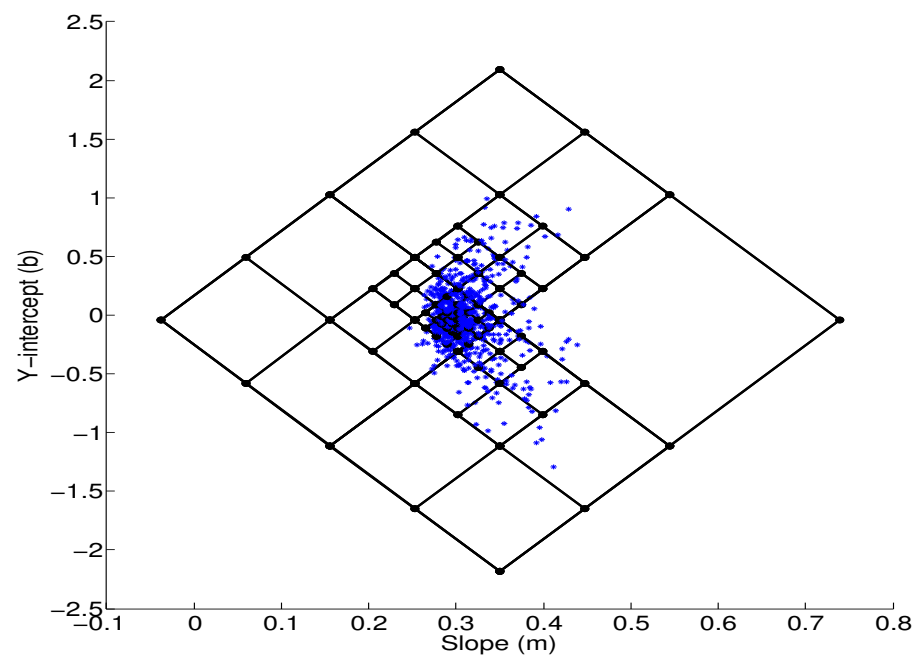

**Figure 7: Quad–tree of bounding regions for a data set. The blue points are the database objects, and the black lines are dominating bounding regions.**

We have now defined the normal lower bound between points in dominance space, as well as a lower bound that is used between a point and a bounding region. We can now group points in dominance space, and prune large amounts of candidate objects with only a single lower bound computation. Computing the distance from $Q$ to a BR $M$ requires finding the minimum/maximum over all the error terms, and with $s$ sub–intervals, is an $O(s)$ operation.

## 5.3 Index Implementation

The index structure implemented is a quad–tree. Quad–trees provide a simple method to define bounding regions, since the regions are based on parallel dominating lines. We can also always ensure that we have a diamond shape to the BR regions by subtracting the mean of the projected bins. This ensures that $t_{min} < 0$ and $t_{max} > 0$, and has no effect on the EMD as it is translation invariant.

Given a data set and a projection $S_j$, we project each distribution onto $S_j$ and fit a 1–dimensional normal distribution to each projection. We pre–compute the errors in each of the $s$ sub–intervals for each distribution. For simplicity, we evenly divide the range from $[t_{min}, t_{max}]$ into $s$ sub–intervals. Finally, we convert each normal distribution to a point in dominance space, determine the bounding region of the entire database, and recursively build the tree. Figure 7 shows an example of the quad–tree for a small sample data set.

At query time, we find potential nearest neighbors by performing a best–first traversal of the quad–tree that utilizes a threshold based on the current $k^{th}$ nearest neighbor. Due to space considerations, we refer the reader to [16] for more details on quad–trees and nearest neighbor querying using a best–first method. The space complexity analysis of the index is discussed in Appendix (C)

Multiple projections may yield a tighter lower bound than a single projection. If more than one projection is used, we implement each projection as its own independent index. We then use a modified version of Fagin's Threshold Algorithm [5] to aggregate the lower bounds from each index.

# 6. EXPERIMENTS

To the best of our knowledge, the TBI method proposed in [21] is the state of the art method for nearest neighbor queries with the EMD. The existing scan–and–refine methods do not implement an index structure, and must perform a large sort of the database lower bounds for each query. Implementation of a lower bound index is more efficient than a single sort on the entire database, as the TBI method demonstrated nearly twice as much improvement in K–NN query time over the previous SAR methods. Therefore, we tested our method only against the TBI method.

As much of the experimental procedure as possible from [21] was replicated in our experiments. The TBI methodology uses an index structure and three additional lower bound filters that are run after candidates are pulled from the index. The filters are run in order: the dual solution lower bound, the reduced EMD lower bound, and the independent minimization lower bound. If a candidate passes all lower bound filters, the full EMD is performed. The implementation of the TBI method was the author's original implementation, with some slight modifications to keep all indices in memory for fair comparison with our method.

The normal lower bound index was coded in C++ and placed into the TBI framework, without the TBI index. The normal lower bound index uses the described index structure(s), then the full projection lower bound, then the reduced EMD lower bound and independent minimization as in the TBI method. All testing was performed on an Intel XEON Quad core processor at 2.33Mhz, with 4GB of RAM.

## 6.1 Data sets

We tested our new method on three real data sets. The first data set is the RETINA data from [11, 20, 21]. This data set consists of 12 MPEG–7 descriptors of 3932 images of the retina. Only the first set of descriptors was used in the experiments. Each descriptor is an $8 \times 12$ grid of tiles. The descriptors were normalized, giving 3932 2–dimensional distributions with 96 total bins. The query objects used for this data set were the same 100 objects used in [20].

The next data set used was the IRMA data set from [20, 21]. This data set consists of features extracted from 10,000 medical images. Each distribution is 199 bins over a 40 dimensional space. The same 100 objects from [20] were used to query the database.

Finally, we used a data set consisting of 681,278 publicly available images from the photo sharing site Flickr. The data set is a collection of general images, each re–sized to be a $640 \times 640$ image. Each image was divided into a $10 \times 10$ grid of tiles, and the 12 feature MPEG–7 color layout descriptor (CLD) was extracted from each tile in an image. Extracting a CLD for each tile achieves a more accurate representation of the spatial distribution of color in an image than using a single CLD for an entire image. The CLDs from each image are then normalized to sum to one. This converts the 100 spatial CLDs to a 2–dimensional probability distribution over the tiles of an image. Each image distribution measures the relative value of the CLD in one tile compared to the CLDs in other tiles of an image. Two images that are close in EMD have similar distribution of CLDs. The CLD distributions model the inherent uncertainty of color features and image comparison. Like the RETINA data, only the first set of descriptors were used. 100 distributions were randomly selected and removed from the data set for querying.

Feasible solutions for the TBI method were generated for all data sets in the random manner as described in [21], and four feasible solutions were employed for each data set, the same as in [21]. For the RETINA and IRMA data sets, the reduction matrices were the same matrices as used in [20]. The reduction lower bound was not used on the Flickr data set due to the generality of the data. That is, reduction matrices provided no pruning power, and actually resulted in a performance decrease due to the overhead of running a reduced EMD. For all tests and data sets, the $L_2$–norm is used as the distance in the cost matrix.

The projection vectors for each data set were found via principal component analysis. As the RETINA and Flickr data sets are two dimensions, we used both principal components as our projection vectors, meaning we utilize two indices. The IRMA data contains almost all variance in one principal component, hence we only employed one projection vector for the IRMA data set. It is possible that there exist better projection vectors to use on these data sets. However, we demonstrate that even using a simple technique such as PCA can yield impressive results.

## 6.2 Results

The only parameters for our tests that we need to vary are the number of sub–intervals, and the node capacity in the indices. We explore how these parameters effect the performance of the method, but we set the default values of these parameters as approximately $log(n)$ number of sub–intervals (5 for RETINA and Flickr, 6 for IRMA) and a node capacity of 100 objects, where $n$ is the number of bins in the data set.

### 6.2.1 K–NN Query Time and Number of EMDs

Figure 8 presents the query time for each method while varying $K$. For the RETINA and IRMA data sets, we present results with the reduction matrix using the original values as presented in [20, 21] (18 and 60, respectively), and with increased values (36 and 80, respectively). As will subsequently be shown, our method significantly decreases the number of full bin EMDs that need to be performed. As a result, we must increase the dimensionality of the reduced filter, or computation time will be wasted on a lower bound filter that has no pruning power.

As one can see, using the original reduction matrix our method reduces the query time. When we take advantage of the available increase in pruning power of the reduction method, the query time for the normal lower bound index decreases even more. As we increase $K$, we maintain the speed up advantage over the TBI method, and on the RETINA and Flickr data sets we see significant improvement as $K$ increases.

On the Flickr data set, the speedup is significant. This speed up is the result of the index's ability to prune many potential candidates before running the other lower bound filters. This can clearly be seen in Figure 9, where we plot the number of full bin EMDs performed. The improvement in the number of full EMDs performed generally follows the pattern of the time speed up. However, on the Flickr data, the reduced lower bound method is useless due to the heterogeneity of the data set. Therefore, the TBI method suffers significant performance degradation without this extra lower bound filter. Our method still successfully prunes many potential candidates without needing the reduced filter, hence we demonstrate its superiority over the TBI method with extremely large data sets.

In addition, the normal index does not require nearly as many data accesses as the TBI method. Recall that the TBI

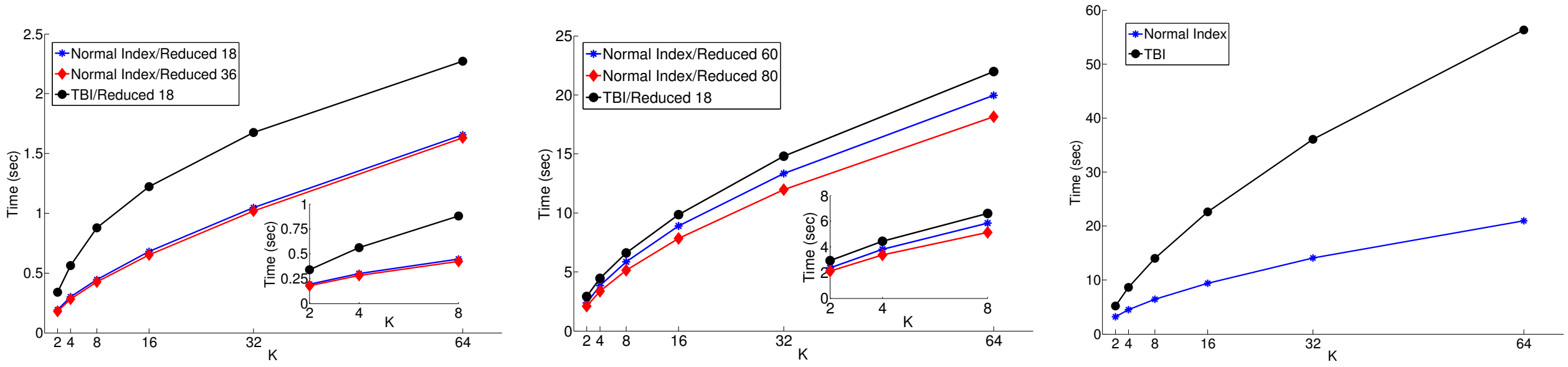

(a) RETINA, 100 Queries (b) IRMA, 100 Queries (c) Flickr, 100 Queries

**Figure 8: Average K–NN query time for all data sets. $K$ values up to 8 are shown in the insets for the RETINA and IRMA data sets. The normal lower bound index outperforms the TBI method on all data sets. The index also enables increased pruning power of the reduced method since the number of full EMDs is drastically reduced. On the Flickr data set, the normal index performance over TBI is significant. Both methods were unable to use the reduced filter, yet the normal index still retains high pruning power.**

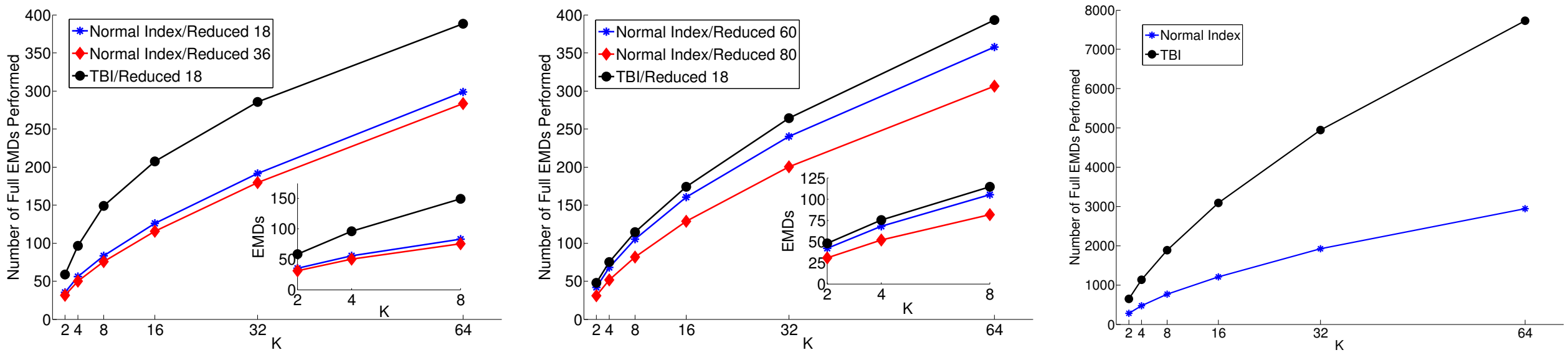

(a) RETINA, 100 Queries (b) IRMA, 100 Queries (c) Flickr, 100 Queries

**Figure 9: Average number of full EMDs performed. The normal index performs very few full bin EMDs, and scales better as $K$ and the database cardinality increases, as demonstrated in the Flickr data set results.**

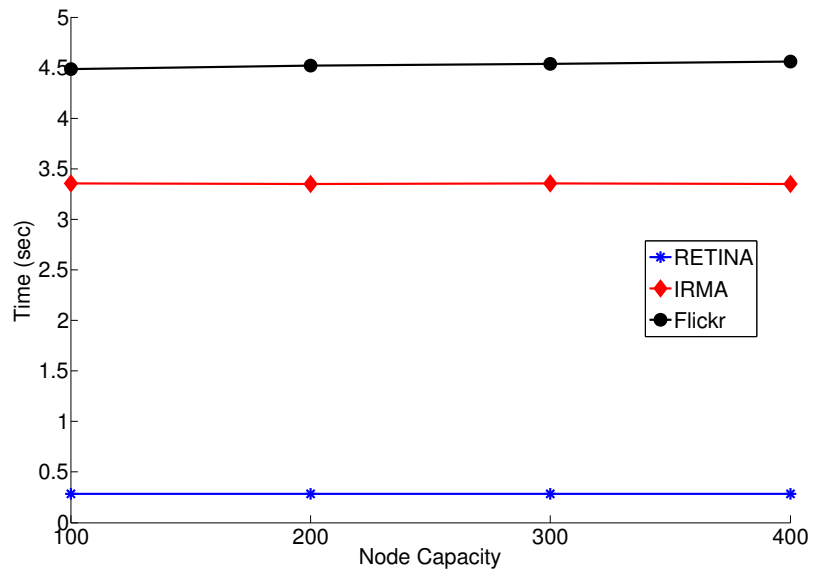

**Figure 10: Average K–NN query time while varying the node capacity. The node capacity has very little impact on query time for any of the data sets.**

method must implement two B+ trees for each feasible solution that is used. Hence, with four feasible solutions, TBI implements eight B+ trees. This results in an extremely high data access rate. The TBI method can access more candidates than the cardinality of the data set, due to the multiple indices accessing the same objects multiple times. The normal lower bound index uses only as many indices as projections, requiring less access to the data than the TBI method. For example, on the Flickr data set, the TBI method accesses nearly four times as much data as the normal index (not shown).

### 6.2.2 *Index Parameters*

We next vary the index parameters to determine what effect they impart on the performance of the method. We vary the number of sub–intervals from 1 to 9, and the node capacity of the index from 100 to 400.

In Figure 10, we examine how changing the internal node capacity in the quad–tree affects the overall performance. As we can see, there is no detectable performance difference. The lower bound computation performed in the index is an $O(s)$ computation for internal nodes, and an $O(1)$ computation for leaf nodes. Therefore, given the large amount of leaf nodes in the index and a low value of $s$, changing the relatively small number of internal nodes does not affect the overall query time significantly.

In Figure 11(a), we vary the number sub–intervals for all three data sets. We see that increasing the number of sub–intervals decreases the query time. On the RETINA and IRMA data sets, there is a slight decrease. On the Flickr data set, the decrease is more significant, as can be seen in more detail in Figure 11(b). In Figure 11(b), as $K$ increases, the difference in query time with more sub–intervals becomes significant. Notice that there are diminishing returns from increasing the number of sub–intervals; as $s$ increases, we get a more accurate lower bound, but eventually it has little impact on the final query time. Finally, we show in Figure 11(c) the number of candidate objects that are eventually passed from the index to the other lower bound filters. As $s$ increases, there is a significant decrease in the number of candidates that are passed to the other lower bound filters. This is expected, as increasing $s$ tightens the normal lower bound in the index. As the database size increases, $s$ can be increased to increase the pruning power of the bound.

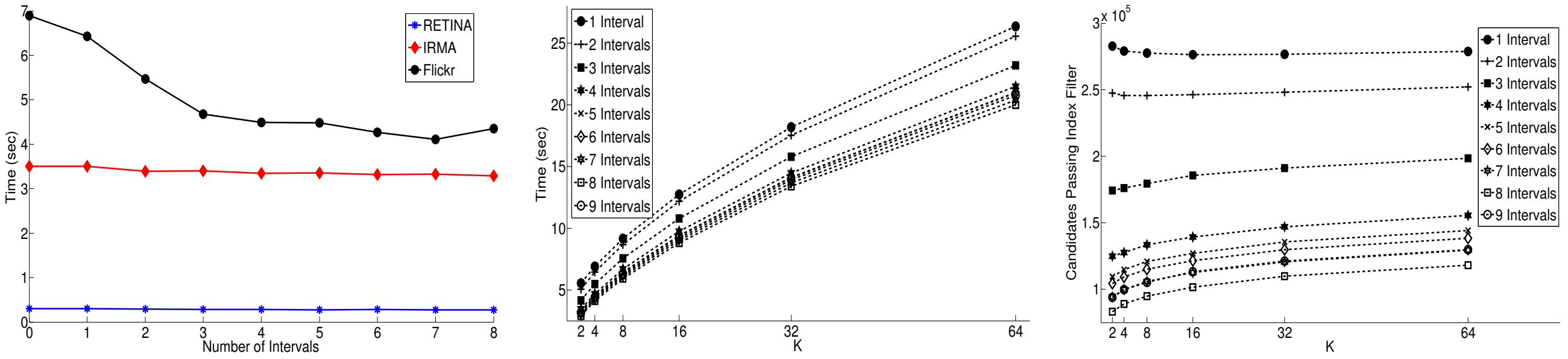

(a) Varying Intervals, K=4 (b) Flickr, Varying K (c) Candidates Passing Index Filter

**Figure 11: Effect of the number of sub–intervals $s$. Figure 11(a) varies the number of sub–intervals for all three data sets with $K = 4$. The number of sub–intervals has a small impact on the RETINA and IRMA data sets, but a large impact on the Flickr data. Figure 11(b) varies $K$ for different values of $s$ on the Flickr data set. When $K$ gets large, the impact of the number of sub–intervals is significant. Figure 11(c) shows the number of candidates that pass the index lower bound for the Flickr data. As $s$ increases, the normal bound in the index becomes tighter and less candidates are passed to the subsequent lower bound filters.**

# 7. CONCLUSION

In this paper, we have presented a novel method for nearest neighbor queries using the Earth Mover's Distance. We introduce a new lower bound to the EMD that approximates the projected EMD using normal distributions and error terms. We develop a novel indexing scheme using stochastic dominance to improve scalability of nearest neighbor queries. We show that our method can retrieve nearest neighbors significantly faster than previous methods.

We anticipate expanding and investigating more aspects of the normal lower bound index. For example, the index requires no use of the projected distribution bins, outside of the minimum and maximum bin values. Therefore, our method is suitable for databases that contain objects without static bins, such as shape databases or moving object databases.

The EMD is a computationally expensive distance metric, but the accuracy and quality it produces in querying and mining makes it highly attractive. As uncertain data sets become larger and more prevalent, extremely high quality and flexible pruning methods will need to be developed to keep query times tractable. We believe our normal lower bound index is a significant step towards achieving these goals.

# 8. ACKNOWLEDGMENTS

We thank Jia Xu for providing the code to the TBI method and Marc Wichterich for providing some of the data sets. This work was supported by NSF grant #0808772.

# APPENDIX

## A. NORMAL LOWER BOUND PROOF

We present the proof to Theorem 2 that

$$EMD_{LB}(\Phi_P, \Phi_Q) \leq EMD(projs_{S_j}(P), projs_{S_j}(Q))$$

Proof. Without loss of generality, we prove the case when $\Phi_P \prec \Phi_Q$. We define $EMD_{LB}(\Phi_P, \Phi_Q)'$ as

$$EMD_{LB}(\Phi_P, \Phi_Q)' = \int_{t_{min}}^{t_{max}} |\Phi_P(t) - \Phi_Q(t)|\, dt - \int_{t_{min}}^{t_{max}} Err_P(t)\, dt + \int_{t_{min}}^{t_{max}} Err_Q(t)\, dt$$

From the definition of the minimum and maximum errors in Eqn. (4), we know that

$$EMD_{LB}(\Phi_P, \Phi_Q) \leq EMD_{LB}(\Phi_P, \Phi_Q)'$$

as $EMD_{LB}(\Phi_P, \Phi_Q)'$ is the computation of the lower bound without pre–computing the intersection errors.

Demonstrating that

$$|\Phi_P(t) - \Phi_Q(t)| - Err_P(t) + Err_Q(t) \leq |C_p(t) - C_q(t)|\ \forall t$$

implies that $EMD_{LB}(\Phi_P, \Phi_Q)'$ is no more than the projection EMD, in which case, we know that $EMD_{LB}(\Phi_P, \Phi_Q)$ is then a lower bound to the projection EMD.

At some point $t$, if $C_p(t) < C_q(t)$, then the contribution to the projected EMD at this point $t$ is

$$\begin{aligned} |C_p(t) - C_q(t)| &= C_q(t) - C_p(t) \\ &= (Err_Q(t) + \Phi_Q(t)) - (Err_P(t) + \Phi_P(t)) \\ &= |\Phi_P(t) - \Phi_Q(t)| - Err_P(t) + Err_Q(t) \end{aligned}$$

Hence the value computed at this point $t$ is the same using the normals and the errors as the projected EMD. If $C_p(t) > C_q(t)$ then the contribution to the EMD is

$$\begin{aligned} |C_p(t) - C_q(t)| &= C_p(t) - C_q(t) \\ &= (Err_P(t) + \Phi_P(t)) - (Err_Q(t) + \Phi_Q(t)) \\ &= (\Phi_P(t) - \Phi_Q(t)) + (Err_P(t) - Err_Q(t)) \\ &\geq |\Phi_P(t) - \Phi_Q(t)| - Err_P(t) + Err_Q(t) \end{aligned}$$

The last line is due to the fact that we assume $\Phi_P \prec \Phi_Q$, so in order for $C_p(t)$ to be greater than $C_q(t)$, $Err_P(t)$ must be greater than $Err_Q(t)$ and the difference between the normals.

Thus for every $t$, the normal EMD plus/minus the error terms from the two distributions will always be at most the actual value contributed to the projected EMD. Hence, $EMD_{LB}(\Phi_P, \Phi_Q)' \leq EMD(projs_{S_j}(P), projs_{S_j}(Q))$, which means that the normal lower bound is also less than the projected bound. The proof for the case where $\Phi_Q \prec \Phi_P$ is the same with the signs reversed. When $\Phi_P(t)$ and $\Phi_Q(t)$ intersect, we can break the proof for the theorem into two parts; before the intersection and after. □

## B. PARTIAL DOMINANCE PROOF

To prove Theorem 5, we need to show that the normal EMD terms and the error terms in the bound are both less than any point in $M$. From the definition of the bound in Eqn. (8), we know that the error terms are always less than any error term in $M$. We now just need to show that the minimum normal EMD between $Q$ and $M$ lies on the line between $M_{is}$ and $M_u$. That is,

$$\underset{m_i \in M}{min}\{EMD_{\mathcal{N}}(\Phi_{m_i}, \Phi_Q)\} \geq \frac{1}{2}[EMD_{\mathcal{N}}(\Phi_{M_u}, \Phi_Q) + EMD_{\mathcal{N}}(\Phi_{M_{is}}, \Phi_Q) - EMD_{\mathcal{N}}(\Phi_{M_{is}}, \Phi_{M_u})]$$

Proof. We can easily demonstrate this via contradiction.

Assume that the minimum EMD normal is a point $W$ not on the line between $M_u$ and $M_{is}$, that is, inside the bounding region. Let us denote the intersection between $\Phi_Q$ and $\Phi_W$ as $t_{is}$. The set of all normal distributions that intersect $\Phi_Q$ at $t_{is}$ lie along a line in dominance space that passes through both $Q$ and $W$. This line can be found by simple arithmetic and we omit the details due to space considerations.

There now exists on the line from $W$ to $Q$, another normal $W'$ with same normal intersection point, but with $\sigma_{w'} > \sigma_w$. Meaning, with a variance that is closer to $\sigma_p$. As $W$ and $W'$ also intersect at $t_{is}$, the CDF difference between $Q$ and $W'$ must be less than $Q$ and $W$ because the intersection points are all the same but $\sigma_{w'}$ is closer to $\sigma_p$ than $\sigma_w$. Hence, on this line from $W$ to $Q$, we can keep finding normals that have a smaller distance to $Q$ than $W$, until we reach the line between $M_u$ and $M_{is}$. Thus, the minimum normal distance must lie on the line between $M_u$ and $M_{is}$. □

## C. SPACE COMPLEXITY ANALYSIS

The projection of each database object is represented by a normal approximation and error terms. This requires storage of the mean and variance of each distribution on the projection, as well as the minimum and maximum errors for all $s$ sub–intervals. Therefore, any index that implements the normal lower bound requires $2NP + 2NPs = O(NPs)$ space, where $N$ is the database cardinality and $P$ is the number of projections. In contrast, the TBI method requires $2LN = O(LN)$ space, where $L$ is the number of feasible solutions that are implemented by the index. As shown in Section 6.2, the number of projections and sub–intervals is generally very small, achieving increased performance over the TBI method with only a modest increase in the space requirements.